\documentclass{article}

\usepackage[final,sglblindworkshop, nonatbib]{neurips_2025} 
\workshoptitle{Data on the Brain \& Mind}

\usepackage{hyperref}
\usepackage{url}             
\usepackage{booktabs}        
\usepackage{amsfonts}        
\usepackage{nicefrac}        
\usepackage{microtype}       
\usepackage{xcolor}          
\usepackage{pifont}          
\usepackage{siunitx}         
\usepackage[normalem]{ulem}  
\usepackage{extarrows}
\usepackage{lipsum}
\usepackage{xspace}

\usepackage{tikz}
\usetikzlibrary{shapes}
\usepackage{dirtytalk}
\usepackage{bbding}         
\usepackage{enumitem}       
\usepackage{empheq}         

\usepackage{algorithm}
\usepackage{algpseudocode}

\usepackage{tabularray}
\usepackage{tabularx}
\usepackage{array}  
\newcolumntype{C}{>{\centering\arraybackslash}X}  

\usepackage{float}
\usepackage{wrapfig}
\usepackage{graphicx}
\graphicspath{{./figs}}

\usepackage{cleveref}
\crefname{figure}{Fig.}{Figs.}
\Crefname{figure}{Figure}{Figures}

\usepackage{caption}

\usepackage{tcolorbox}
\usepackage[makeroom]{cancel}

\usepackage[utf8]{inputenc}     
\usepackage[T1]{fontenc}        

\usepackage{amsmath, amssymb, amsfonts, mathtools}
\usepackage{bm}

\usepackage{todonotes}

\DeclareRobustCommand{\xx}{\bm{x}}

\DeclareRobustCommand{\zz}{\bm{z}}

\DeclareRobustCommand{\blamb}{\bm{\lambda}}

\DeclareMathOperator{\EE}{\mathbb{E}}

\DeclareMathOperator{\PP}{\mathcal{P}}

\newcommand{\norm}[1]{\left\lVert#1\right\rVert}

\newcommand{\expect}[2]{\EE_{#1}\!\Big{[} {#2} \Big{]}}

\def\1{\bm{1}}

\def\rva{{\mathbf{a}}}

\def\rvr{{\mathbf{r}}}

\def\rvx{{\mathbf{x}}}

\def\rvz{{\mathbf{z}}}

\DeclareMathAlphabet{\mathsfit}{\encodingdefault}{\sfdefault}{m}{sl}
\SetMathAlphabet{\mathsfit}{bold}{\encodingdefault}{\sfdefault}{bx}{n}

\definecolor{neuro}{HTML}{39A300}
\definecolor{comp}{HTML}{077AF8}
\definecolor{theo}{HTML}{EF4136}
\definecolor{RowColorTop}{HTML}{e0dccb}
\definecolor{RowColorDark}{HTML}{f2efe6}
\definecolor{RowColorLight}{HTML}{fdfbf7}
\definecolor{MSBlue}{rgb}{.204,.353,.541}
\definecolor{MSLightBlue}{rgb}{.31,.506,.741}
\definecolor{BlogBlue}{HTML}{2d6a99}
\definecolor{BlogBG}{HTML}{f9f6f5}
\definecolor{py_0}{HTML}{1F77B4}
\definecolor{py_1}{HTML}{FF7F0E}
\definecolor{py_2}{HTML}{2CA02C}
\definecolor{py_3}{HTML}{D62728}
\definecolor{py_4}{HTML}{9467BD}
\definecolor{py_5}{HTML}{8C564B}
\definecolor{py_6}{HTML}{E377C2}
\definecolor{py_7}{HTML}{7F7F7F}
\definecolor{py_8}{HTML}{BCBD22}
\definecolor{py_9}{HTML}{17BECF}
\definecolor{gr}{HTML}{37C532}
\definecolor{grey}{HTML}{4D4D4D}
\definecolor{gray}{HTML}{4D4D4D}
\definecolor{lightgray}{HTML}{666666}
\definecolor{lightlightgray}{HTML}{999999}
\definecolor{color_enc}{HTML}{ED1C24}
\definecolor{color_dec}{HTML}{29ABE2}
\definecolor{color_decenc}{HTML}{dd1aff}

\newcommand{\pois}{\mathcal{P}\mathrm{ois}}
\newcommand{\pvae}{$\PP$-VAE\xspace}

\newcommand{\underbracegray}[2]{
    {\color{lightlightgray}
        \underbrace{\color{black}#1}_{\color{black}#2}
    }
}

\NewDocumentCommand{\boxit}{ O{} m }{
    \begin{tcolorbox}[
        colback=yellow!20, 
        colframe=lightgray, 
        title={#1} 
    ]
    \centering
    {#2} 
    \end{tcolorbox}
}

\usepackage[
    backend=biber,
    sorting=none,
    style=numeric-comp,
    maxnames=2,
    minnames=1
]{biblatex}
\usepackage[misc]{ifsym}
\usepackage{enumitem}
\usepackage{mdframed}
\usepackage{url}  

\renewcommand{\url}[1]{} 

\title{Inferring response times of perceptual decisions with Poisson variational autoencoders}

\makeatletter
\newcommand\blfootnote[1]{%
  \begingroup
    \renewcommand\thefootnote{}%
    \begin{NoHyper}%
      \footnote{#1}%
    \end{NoHyper}%
    \addtocounter{footnote}{-1}%
  \endgroup
}
\makeatother

\author{%
\textbf{Hayden R. Johnson}$^{1,2}$ \quad
\textbf{Anastasia N. Krouglova}$^{1,2}$ \quad
\textbf{Hadi Vafaii}$^{4}$ \\[0.3em]
\textbf{Jacob L. Yates}$^{3,4\;\dagger}$ \quad
\textbf{Pedro J. Gon\c{c}alves}$^{1,2\;\dagger}$\\[0.45em]
$^1$VIB–Neuroelectronics Research Flanders, Belgium \quad
$^2$KU Leuven, Belgium\\[0.1em]$^3$Herbert Wertheim School of Optometry \& Vision Science, UC Berkeley, USA\\[0.1em]
$^4$Redwood Center for Theoretical Neuroscience, UC Berkeley, USA
\\[0.25em]
\texttt{hayden.johnson@vib.be}
}

\begin{document}

\maketitle

\begin{abstract}
Many properties of perceptual decision making are well-modeled by deep neural networks.  However, such architectures typically treat decisions as instantaneous readouts, overlooking the temporal dynamics of the decision process. We present an image-computable model of perceptual decision making in which choices and response times arise from efficient sensory encoding and Bayesian decoding of neural spiking activity. We use a Poisson variational autoencoder to learn unsupervised representations of visual stimuli in a population of rate-coded neurons, modeled as independent homogeneous Poisson processes. A task-optimized decoder then continually infers an approximate posterior over actions conditioned on incoming spiking activity. Combining these components with an entropy-based stopping rule yields a principled and image-computable model of perceptual decisions capable of generating trial-by-trial patterns of choices and response times. Applied to MNIST digit classification, the model reproduces key empirical signatures of perceptual decision making, including stochastic variability, right-skewed response time distributions, logarithmic scaling of response times with the number of alternatives (Hick's law), and speed–accuracy trade-offs.\blfootnote{${\dagger}$\; Co-senior authors.}

\end{abstract}

\section{Introduction}
A central task of the brain is to process sensory information in order to guide adaptive behavior, often studied through perceptual decision-making. This process is shaped by several biological constraints. Neurons communicate through discrete action potentials (or spikes), the nervous system is subject to noise at multiple levels \cite{faisal2008noise}, and operates under limited metabolic resources \cite{laughlin2001energy, stone2018principles}. Together, these factors impose fundamental limits on both the speed and accuracy of perceptual processing \cite{heitz2014speed}.

Deep neural networks have emerged as powerful models of the sensory cortex, capable of performing perceptual tasks of naturalistic complexity \cite{yamins2014performance, kell2018task}. While compelling, these models typically treat decisions as instantaneous readouts, ignoring the rich temporal dynamics of the decision process. As a result, standard deep neural networks cannot account for temporal properties of perceptual decision making such as response time distributions or the speed–accuracy trade-offs.

In contrast, evidence accumulation models (e.g., drift diffusion model \cite{ratcliff1978theory}) explicitly capture the temporal dynamics of the decision process. These models reproduce key behavioral regularities, including stochastic variability, right-skewed response time distributions, and speed–accuracy trade-offs \cite{evans2020evidence}. However, they generally operate on abstract decision variables without specifying how such variables are derived from sensory input. Consequently, they are not image-computable (i.e., they do not compute directly on visual stimuli) and therefore cannot perform image-level perceptual tasks.

Recently, a new class of image-computable decision models has sought to fill this gap by explicitly linking sensory encoding with decision dynamics \cite{chen2017seeing, goetschalckx2023computing, cheng2024rtify, rafiei2024neural, jaffe2025image}. These models seek to combine the representational capacity of deep neural networks with temporal accumulation of evidence, providing a functional grounding for the latent decision variable in terms of task-relevant information. Doing so enables models to perform complex tasks while accounting for temporal properties of the decision process. However, many existing approaches are directly fit to human response data, or introduce temporal integration mechanisms without a clear normative justification.

In this work, we introduce \pvae-RT, an image-computable model for perceptual decision-making, grounded in principles of efficient coding and Bayesian evidence accumulation. Sensory representations are modeled as resource-limited encoders optimized for information transmission in a population of rate-coded neurons. Decisions are then cast as Bayesian accumulation of noisy sensory evidence (via spikes) until a criterion is reached. This formulation allows the model to perform image-level tasks while reproducing key behavioral phenomena observed in perceptual decision making. By unifying efficient sensory coding with normative accumulation of evidence, our approach provides a framework for understanding perceptual decision-making as optimal inference under resource constraints \cite{bhui2021resource}.

\textbf{Contributions: } Overall, our work makes the following contributions: 
\textbf{(i)} We introduce an image-computable model of perceptual decision-making that integrates efficient sensory encoding with Bayesian evidence accumulation, providing a principled link between high-dimensional sensory input and temporal decision dynamics. 
\textbf{(ii)} We demonstrate our model's ability to perform a complex image-level task while predicting hallmark behavioral phenomena, including trial-to-trial variability, response time distributions, and adaptive speed–accuracy trade-offs.

\begin{figure*}[t!]
\begin{center}
\centerline{\includegraphics[width=\linewidth]{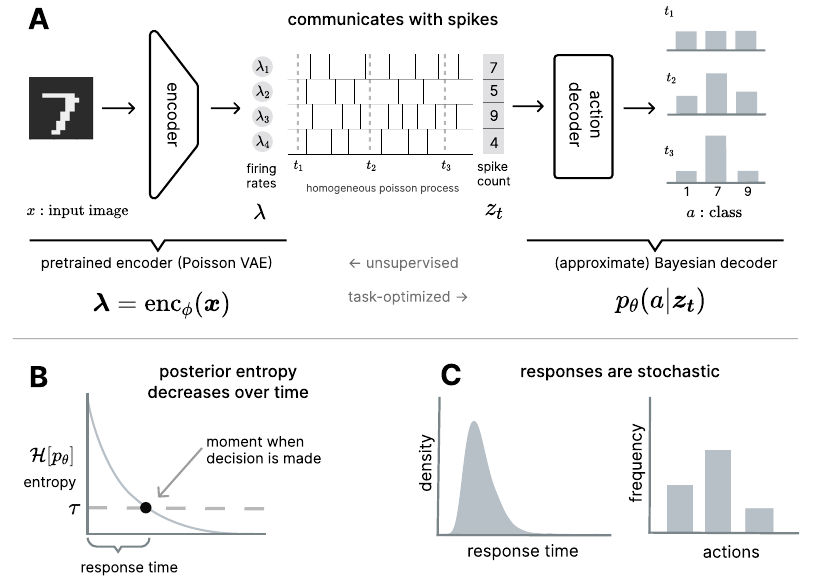}}
\caption{\textbf{(A)} \pvae-RT architecture. Input stimuli $\rvx$ are processed by a pretrained \pvae encoder, $\text{enc}_\phi(\rvx)$, producing a vector of firing rates $\boldsymbol{\lambda}$. These rates generate spike trains via a set of homogeneous Poisson processes. Throughout the spike train, an approximate Bayesian decoder continually infers the posterior distribution, $p_\theta(\rva \mid \rvz_t)$, over actions $\rva$ based on the accumulated spike count $\rvz_t$. \textbf{(B)} Schematic of the entropy-based stopping rule. Posterior entropy $\mathcal{H}[p_\theta]$ decreases as spikes accumulate. Response times are modeled as the first passage time for the posterior to hit an entropy stopping threshold $\tau$. \textbf{(C)} Schematic of response distributions. We generate response distributions from repeated simulation of actions and response times for a given stimulus.
\label{fig:method}
}
\end{center}
\end{figure*}
\section{Background}

\textbf{Bayesian neural decoding.} 
Previous work demonstrated that several classic response-time phenomena can be unified from the perspective of information transmission. In particular, \textcite{christie2023information} considers response times (RT) as the duration required for a Bayesian decoder to identify a stimulus from a population of rate-coded neurons, each modeled as a homogeneous Poisson process.

Formally, consider a set of stimuli $\mathcal{X} =\{x_1,..,x_K\}$ where each stimulus $
x \in \mathcal{X}$ is sampled from a prior distribution $p(x)$. The stimulus is encoded into vector firing rates 
$$\boldsymbol{\lambda} = \rho_b + \gamma \; g(x), \qquad  g: \mathcal{X} \rightarrow \mathbb{R}^K, $$
where $g$ is a one-hot encoding, $\gamma$ scales the encoding magnitude, and $\rho_b$ is a uniform baseline firing rate. In other words, each stimulus excites a unique neuron, while all others remain at baseline activity. Spikes are then generated according to a set of homogeneous Poisson processes with these rates. The decoder is modeled as a Bayesian ideal observer that infers the stimulus from cumulative spike counts at time $t$, denoted $\rvz_t$, by updating the posterior
$$p(x \mid \rvz_t) = \frac{p(\rvz_t \mid x) \; p(x)}{p(\rvz_t)}.$$
Under this simple one-to-one encoding, the posterior can be computed in closed-form \cite{christie2019information}. Decoder uncertainty is quantified by the entropy $\mathcal{H}(x \mid \rvz_t)$, which decreases as spikes accumulate. Response time is defined as the first-passage time when entropy drops below a decision threshold $\tau$.

Despite its simplicity, this framework provides a unified account for a wide range of empirical findings, including the Hick–Hyman law \cite{hick1952rate, hyman1953stimulus}, the power law of practice \cite{newell2013mechanisms}, Stroop interference \cite{macleod1991half, stroop1935studies}, and speed–accuracy trade-offs \cite{wickelgren1977speed, heitz2014speed}. However, extending the model to complex tasks reveals two key limitations. First, the model assumes a one-to-one neuron–stimulus mapping, incommensurate with complex encoders required for high-dimensional stimuli (e.g., images). Second, the model fails to accommodate non-identity relationships between stimuli and actions, preventing it from capturing tasks with complex or continuous outputs. In this work, we address these limitations by generalizing the framework to support complex, nonlinear stimulus encoders and by developing a task-optimized decoder capable of supporting complex actions.

\textbf{Poisson variational autoencoder.}
Perception can be understood as the process of inferring the true state of the world from noisy and incomplete sensory measurements \cite{von1867handbuch}. This perspective has directly inspired machine learning architectures such as the Helmholtz machine \cite{dayan1995helmholtz}, and, more recently, variational autoencoders (VAEs; \cite{kingma2013auto,rezende2014stochastic}).

VAEs are increasingly being considered as computational models of perceptual inference in the brain. Evidence for this comes from several directions: (i) VAE representations align with the primate cortex across both ventral \cite{higgins2021unsupervised} and dorsal \cite{vafaii2023hierarchical} streams; (ii) VAEs develop cortex-like topographic organization \cite{keller2021modeling, keller2021topographic}; and (iii) VAEs produce perceptual errors resembling those of humans \cite{storrs2021unsupervised}. Collectively, this suggests a significant degree of neural, organizational, and psychophysical alignment between VAEs and the brain.

However, standard Gaussian VAEs encode inputs into continuous latent variables that deviate from the discrete, spiking nature of biological neural codes. To address this, \textcite{vafaii2024poisson} introduced the Poisson VAE (\pvae), which encodes inputs into discrete spike-count variables. The \pvae is both rigorously grounded in the mathematics of variational inference, and is more brain-like in its representations.

Despite these advances, the \pvae remains a purely vision-based model. It has not yet been applied to scenarios where perceptual inference is used to guide behavior. In this work, we extend the \pvae framework to demonstrate that its encoding mechanism can generate spike trains that support perceptual decision-making.

\section{Overview of the \pvae-RT}
We construct our model (\pvae-RT) from two components, each trained independently. First, we train a Poisson variational autoencoder to learn an efficient unsupervised representation of stimuli as firing rates. These rates are then converted into spike trains via a set of homogeneous Poisson processes. The spike trains are then used as training data for a task-optimized decoder that learns to continually infer the posterior over actions, conditioned on incoming spiking activity. Combining these components with an entropy-based stopping rule yields a principled, image-computable model of perceptual decision-making that generates both choices and response times. In the following sections, we describe each component in detail and outline its role within the unified architecture.

\subsection{Efficient, probabilistic encoding of stimuli}
To efficiently encode visual stimuli into a vector of firing rates, we use a Poisson VAE (\pvae). The \pvae modifies a standard variational autoencoder by replacing the Gaussian prior and approximate posterior with Poisson distributions. This modification results in discrete, integer spike-count representations in the latent space of the model. 

During inference, the \pvae encoder maps an input sample $\xx$ to the rate parameters, $\blamb(\xx) = \mathrm{enc}_\phi(\xx)$, which are then used to construct a Poisson approximate posterior, from which spike counts $\zz \sim \mathrm{Pois}(\blamb(\xx))$ are sampled. The decoder network then reconstructs the input, $\hat{\xx} = \mathrm{dec}_\psi(\zz)$.

The \pvae is trained in an unsupervised manner, using the standard \underline{E}vidence \underline{L}ower \underline{BO}und (ELBO) objective, which assumes the following form for a Gaussian likelihood and Poisson latents:
\begin{equation}\label{eq:pvae-elbo}
    \mathcal{L}_{\mathrm{PVAE}}
    \;=\;
    \underbracegray{
        \expect{\zz \sim \pois(\zz; \blamb(\xx))}{\norm{\xx - \mathrm{dec_{\psi}}(\zz)}_2^2}
    }{\text{Reconstruction term}}
    \;+\;
    \underbracegray{
        \sum_{i=1}^{K} r_i f(\delta r_i)\
    }{\text{KL term}},
\end{equation}
where $\mathrm{dec}_\psi(\cdot)$ denotes the decoder network, $K$ is the latent dimensionality, and $f(y)=1-y+y\log y$; $\rvr$ is the prior over firing rates, and $\delta\rvr \coloneqq \boldsymbol{\lambda}\oslash\rvr$, where $\oslash$ denotes element-wise division.

A notable consequence of incorporating Poisson latents is that the Kullback–Leibler (KL) term in the ELBO objective resembles a metabolic cost (\cref{eq:pvae-elbo}) \cite{vafaii2024poisson}. Because of this mathematical result, the overall \pvae objective encourages the model to faithfully reconstruct inputs while minimizing spiking activity. This establishes a theoretical connection to sparse coding \cite{olshausen1996emergence}, which was verified empirically \cite{vafaii2024poisson, vafaii2025ipvae}.

In sum, the \pvae provides a biologically plausible, spiking, and sparsity-promoting perceptual model, which we use as a \say{visual cortex} in our decision-making task.

\subsection{Task-optimized neural decoding} 
Following \textcite{christie2023information}, we model response time as the time required for a Bayesian decoder to infer sensory information from spikes emitted by a population of rate-coded neurons. We extend this framework by allowing the decoder to interpret non-linear encodings and learn arbitrary mappings between stimuli and actions.  


Concretely, we consider the problem of inferring the posterior
$p(a \mid \zz_t) \propto p(\zz_t \mid a) \; p(a)$, where $a$ denotes a perceptual judgment (e.g., a class label) pertaining to stimulus $\xx \in \mathbb{R}^D$. $\zz_t \in \mathbb{N}^K$ is the vector of cumulative spike counts up to time $t$ from $K$ homogeneous Poisson neurons with rates
$$\blamb = \mathrm{enc}_\phi(\xx), \qquad \mathrm{enc}_\phi(\cdot): \mathbb{R}^D \rightarrow\mathbb{R}^K.$$
We place no restrictions on the complexity of the encoder $\mathrm{enc}(\cdot)$, or the relation between $\xx$ and $a$. This generalization renders the likelihood $p(\zz_t \mid a)$ intractable. We therefore seek to directly approximate the posterior with a learned decoder:
\begin{equation}
     \text{dec}_\theta := p_\theta(a \mid \zz_t) \approx p(a \mid \zz_t),
\end{equation}
where $\theta$ are the neural network parameters that parameterize our approximate posterior. We emphasize that $\text{dec}_\theta$ is completely distinct from the decoder used to train our \pvae (which learns to reconstruct the original stimuli). Rather, this network computes an evolving posterior over actions which may be a complex function of the original stimuli (e.g., the class of an image).

To train this decoder, we construct a dataset $\mathcal{D} = {(\zz_t, a)}$. For each stimulus, we (i) generate spike trains from the image-driven rates $\blamb$ using a homogeneous Poisson process, (ii) discretize the spike train into time bins to form a binary event matrix of size $K \times T$, and (iii) convert this into cumulative spike counts, where entry $(i,t)$ is the total number of spikes emitted by neuron $i$ up to time $t$. Each column of this cumulative count matrix, paired with its label $a$, serves as a training sample. Importantly, for a homogeneous Poisson process, the likelihood depends only on the number of events in an interval, not their precise timing. Thus, the cumulative spike count vector $\zz_t$ is a sufficient statistic for posterior inference \cite{kingman1992poisson}.

In the case of a discrete action space, as seen in the subsequent example, we use a single multi-layer perceptron trained with a softmax output and cross-entropy loss:
\begin{equation}
    \mathcal{L}(\theta) = \mathbb{E}_{(a, \rvz_t)}[-\text{log}\;p_\theta(a \mid \rvz_t)].
\end{equation}
In appendix \ref{app:proof-cond-mle}, we provide a proof that this objective converges to the true posterior under standard realizability assumptions. For continuous actions, we replace the softmax with a
parametric density (e.g., Gaussian mixture model, normalizing flow) and train the decoder using the negative log-likelihood loss.

\subsection{Generating responses from \pvae-RT}
To construct \pvae-RT, we combine a pretrained encoder $\text{enc}_\phi(x)$, which maps high-dimensional stimuli into firing rates, with a task-optimized decoder $p_\theta(a \mid \rvz_t)$, that continually extracts task-relevant information from resulting spike trains (see Figure \ref{fig:method}). 

A simulation proceeds as follows: the input stimulus $\xx$ is first encoded into firing rates $\boldsymbol{\lambda} = \text{enc}_\phi(\xx)$. Spike trains are then generated from a set of homogeneous Poisson processes parameterized by these rates. At each time $t \in [0,T]$, the decoder infers the posterior distribution $p_\theta(a \mid \rvz_t)$ and its associated entropy $\mathcal{H}[p_\theta]$. Response time is defined as the earliest $t$ for which $\mathcal{H}[p_\theta] < \tau$, where $\tau$ is a fixed entropy threshold. The selected action is given by the maximum a posteriori (MAP) estimate of the decoder at that time.

It is worth noting that the model is not trained to mimic subjects' empirically observed behavior. Instead, it is trained directly to perform a task, subject to a few guiding principles: (i) stimuli should be encoded efficiently, subject to representational constraints, (ii) information should be transmitted in a biologically plausible manner (i.e., using spikes), and (iii) task-relevant information should be decoded optimally to support action. Together, these guiding principles provide a powerful framework for understanding how temporal properties of perceptual decision arise in neural systems under biological constraints.

\begin{figure*}[ht!]
\begin{center}
    \centerline{\includegraphics[width=\linewidth]{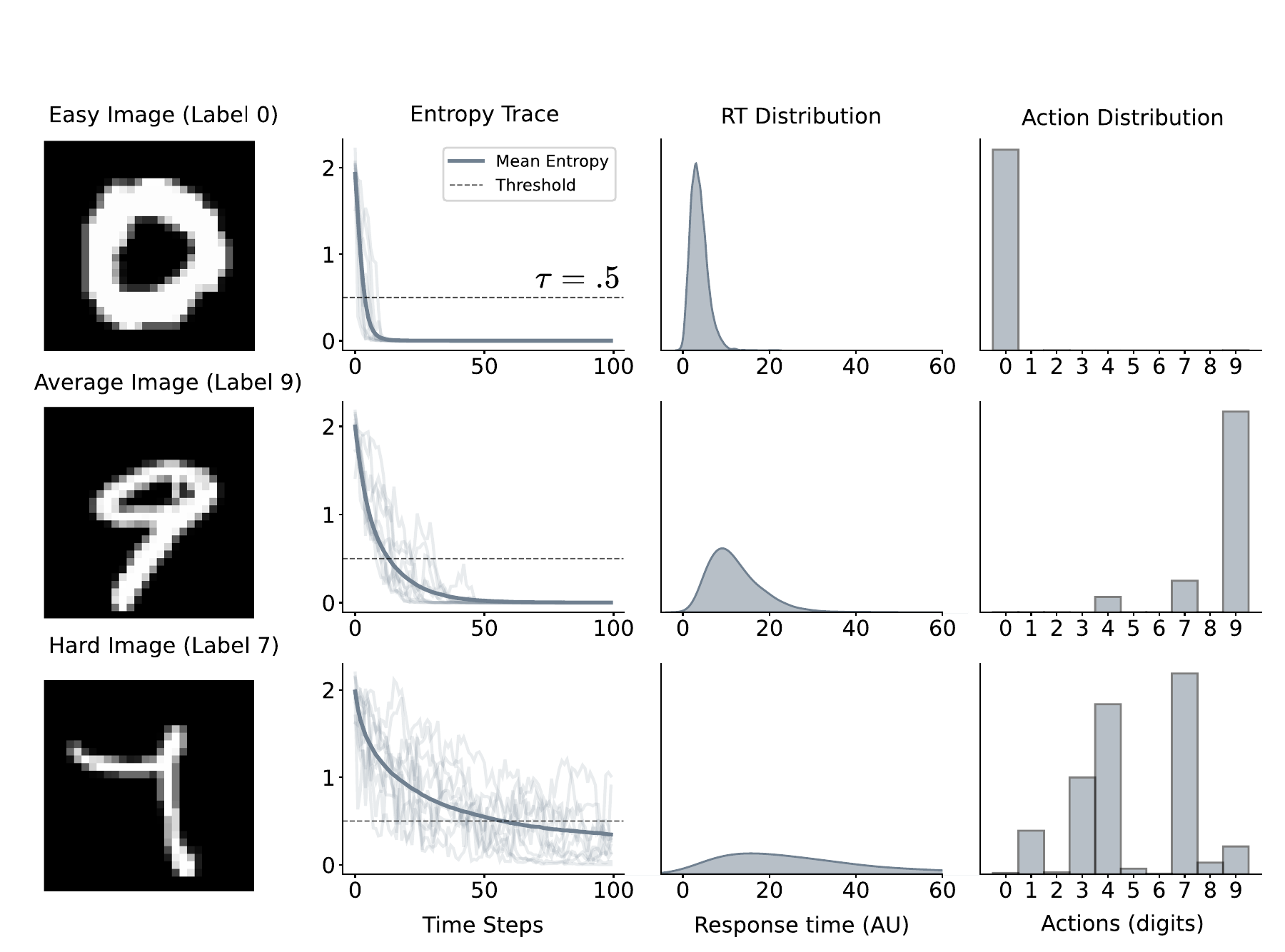}}
\caption{
    \textbf{Response distributions depend on task difficulty.}
    Response distributions for stimuli of varying difficulty. Each row corresponds to a digit at a given difficulty level, while each column highlights a property of the response distribution across repeated trials with a fixed stimulus. Easy stimuli are characterized by rapidly decreasing entropy, strongly skewed distributions, and low variance in the action distribution. Difficult stimuli, by contrast, exhibit slower entropy reduction, more symmetric distributions, and greater variability across actions.
    }
\label{fig:response-distribution}
\end{center}
\end{figure*}

\section{Capturing psychophysical phenomena with MNIST}
We apply \pvae-RT to the task of handwritten digit classification using the MNIST dataset and show that it captures several hallmark regularities of human perceptual decision-making. While this serves as a concrete test case, it also illustrates the model’s capacity to perform a general class of perceptual tasks. 

To evaluate behavior, we report response times in arbitrary units (AU). These can be mapped to real time (e.g., seconds) by specifying a time constant, which in practice can be estimated from empirical data. For the purposes of this section, however, we emphasize the model’s ability to reproduce well-established psychophysical trends, which remain invariant under rescaling of the time axis.

Unless otherwise noted, we use an entropy stopping threshold of $\tau = 0.5$ (except when systematically varied to study the speed–accuracy trade-off) and a latent dimension of 128. Additional analyses on the effect of varying the latent dimensions are provided in appendix~\ref{app:varying-latent} and further details on the training procedure can be found in Appendix \ref{app:training}.

\textbf{Response distributions:} 
Human response times are among the most extensively studied measures in behavioral psychology, providing a unique window into the processes that underlie perceptual decision-making. A key reason is that response time distributions, together with behavioral outcomes, reliably exhibit several regularities. We highlight three core phenomena and show that our model reproduces them: (i) decisions are stochastic, i.e., presenting the same stimulus can yield different responses and response times \cite{beck2012not, renart2014variability}; (ii) response time distributions are typically right-skewed; and (iii) the degree of skewness decreases as task difficulty increases \cite{forstmann2016sequential, evans2020evidence}.

Our model successfully reproduces all three effects (Figure~\ref{fig:response-distribution}). To illustrate this, we selected three images that differ in classification difficulty for human observers. The easy image is quickly and reliably classified as a 0, producing a strongly right-skewed response-time distribution with minimal variance across actions. The average difficulty image is labeled as a 9, but its white pixels also resemble a 4 or a 7, leading to slightly slower response times, reduced skew in the response-time distribution, and increased variance across actions. The hard image closely resembles both a 4 and a 7, creating ambiguity that generates a bimodal action distribution, substantially slower response times, and a broad response time distribution.
\begin{figure*}[hb!]
\begin{center}\centerline{\includegraphics[width=\linewidth]{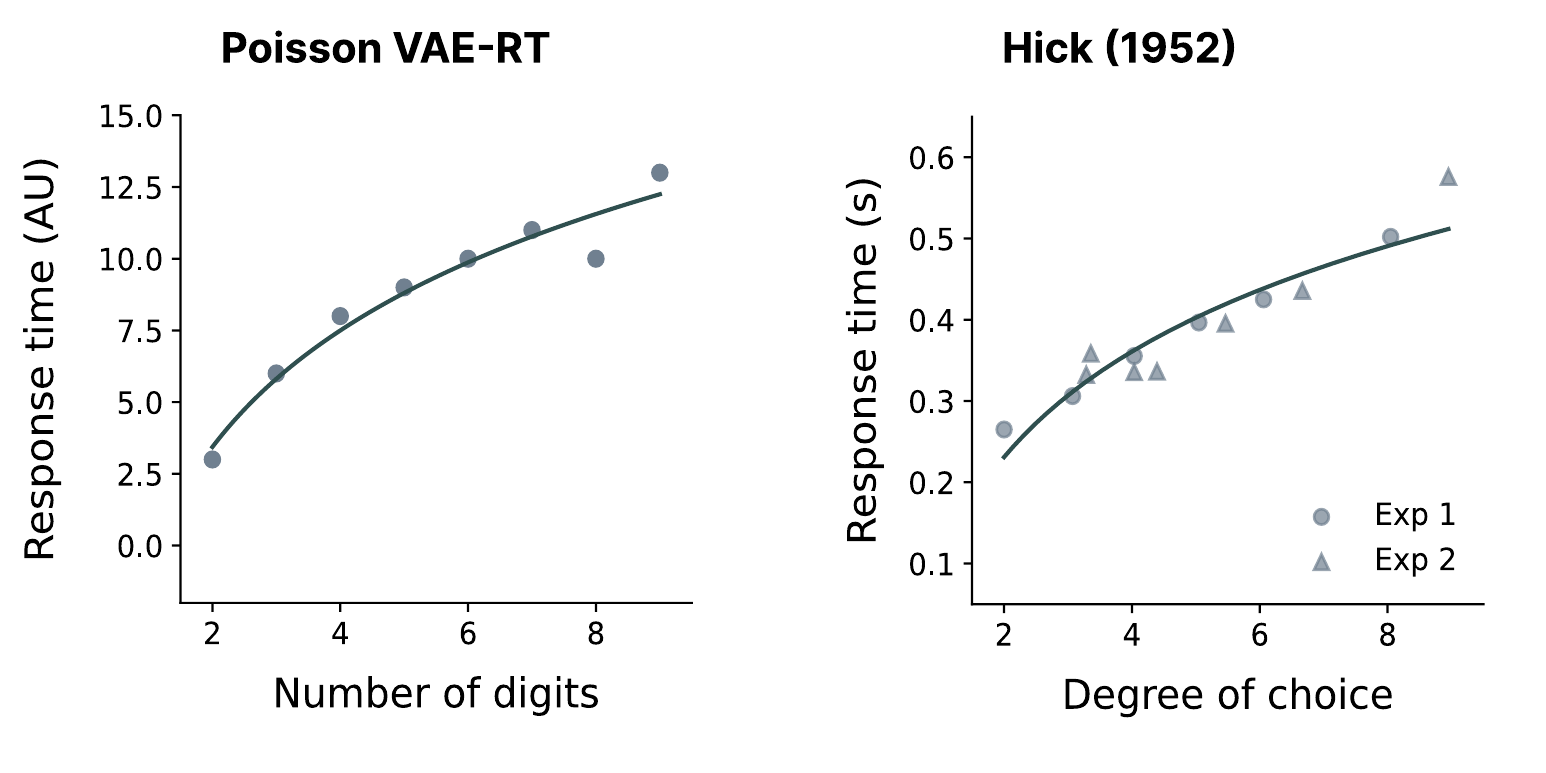}}
    \caption{
        \textbf{Hick's Law}.  (\textit{Left}) Mean response time from the \pvae-RT where the decoder is trained to classify among a varying number of MNIST digits. RT increases monotonically with the number of alternatives with an approximately logarithmic trend. (\textit{Right}) Human response times replotted from Hick \cite{hick1952rate}, shown in seconds.
        }
    \label{fig:hick}
\end{center}
\end{figure*}

\begin{figure*}[hb!]
\begin{center}
    \centerline{\includegraphics[width=\linewidth]{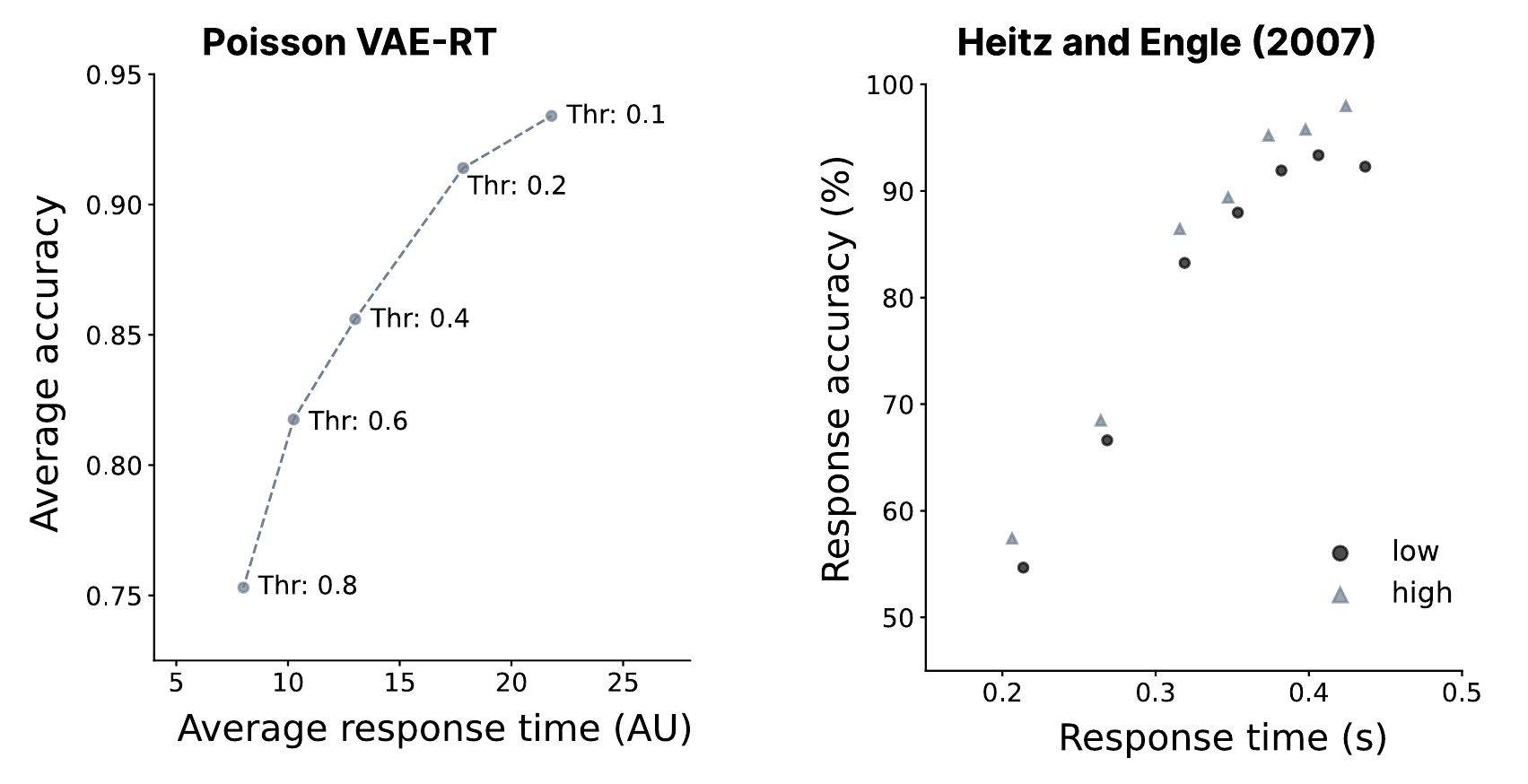}}
    \caption{\textbf{Speed-accuracy trade-off.}
    (\textit{Left}) Model accuracy versus average RT as the entropy threshold is swept ($\tau \in \{0.1,\,0.2,\,0.4,\,0.6,\,0.8\}$).
    Lower $\tau$ values produce longer RTs and higher accuracy. Points represent averages across images and trials. (\textit{Right}) Human response time data from a working memory task under low versus high time pressure (Heitz \& Engle \cite{heitz2007focusing}). 
    }
    \label{fig:speed-accuracy}
\end{center}
\end{figure*}

\textbf{Reproducing the Hick–Hyman law.}
One of the earliest applications of information theory to human behavior is the Hick–Hyman law, which states that response time increases linearly with the information content of a stimulus. Given a uniform prior over stimuli, this implies a logarithmic relationship between response time and the number of possible choices. This effect was first demonstrated by William Hick in a classic experiment where subjects responded to a variable number of stimuli \cite{hick1952rate}, and further developed by Ray Hyman \cite{hyman1953stimulus}. Together, these studies suggest that response times scale with stimulus uncertainty, consistent with the idea that human decision-making proceeds at an approximately constant rate of information processing.

Our model reproduces this trend, when the decoder is trained to classify among varying numbers of digit classes. As shown in Figure~\ref{fig:hick}, response times increase systematically with the number of alternatives, in agreement with the classic behavioral data of \textcite{hick1952rate}.

\newpage
\textbf{Speed-accuracy trade-off: }Biological organisms must often act under variable time constraints, balancing the need for rapid responses against the benefits of accuracy. This balance is formalized by the speed–accuracy trade-off (SAT), whereby faster decisions typically yield higher error rates, while slower decisions improve accuracy at the cost of timeliness. The phenomenon has been documented across a wide range of species and tasks, including perceptual and cognitive decisions in humans \cite{bogacz2010humans, heitz2014speed, drugowitsch2015tuning}, macaques \cite{hanks2014neural}, rats \cite{mendoncca2020impact}, and even insects such as bees \cite{chittka2003bees}. 

Our model reproduces this trade-off, when the entropy stopping threshold $\tau$ is modulated. As illustrated in Figure~\ref{fig:speed-accuracy}, higher thresholds yield faster responses at the cost of accuracy, whereas lower thresholds prolong evidence accumulation and improve performance. This mirrors behavioral trends observed in working-memory experiments with varying levels of time pressure \cite{heitz2007focusing}. By linking stopping rules to decision performance, our framework provides a principled account of how organisms can tune their behavior to the demands of the environment.

\section{Discussion}
In this work, we introduced \pvae-RT, an image-computable model of perceptual decision-making grounded in the principles of efficient coding and Bayesian evidence accumulation. The model employs an unsupervised representation from a Poisson variational autoencoder to map high-dimensional stimuli into a population of rate-coded neurons, modeled as a set of independent homogeneous Poisson processes. These rates generate spike trains that drive a task-optimized decoder that computes an approximate Bayesian posterior over actions throughout the spike train. As spikes accumulate, posterior entropy decreases until reaching a preset threshold $\tau$, at which point the model commits to an action and reports a response time.

We demonstrated the framework on the task of MNIST digit classification and showed that it reproduces classic psychophysical regularities. These include response variability, right-skewed response-time distributions, the logarithmic increase in response times with the number of alternatives (Hick’s law), and speed–accuracy trade-offs.

We believe this approach makes several contributions to the study of perceptual decision-making. From a pragmatic perspective, it offers a highly flexible framework for modeling perceptual decisions with minimal constraints on stimulus complexity or the space of potential actions. We see this flexibility as an important step toward investigating decision-making in more complex and naturalistic contexts. At a conceptual level, our model provides a principled link between efficient sensory coding and evidence accumulation, highlighting how the variability and resource limitations of neural population activity can give rise to variability in behavior. From a broader perspective, this work underscores the value of response times as a meaningful axis for comparing artificial and biological neural networks, extending alignment efforts beyond static measures such as accuracy or representational similarity.

\textbf{Limitations and future work.} While we view these results as promising, several limitations remain to be addressed in future work. First, we have not yet performed a direct quantitative comparison to human behavioral data, an essential step in assessing the model’s explanatory power. Applying the framework to empirical datasets requires estimating several free parameters, including the time constant, entropy threshold, and neural population size. Future work should explore how these parameters can be constrained by normative principles or efficiently fit to behavioral measurements.

Second, the current formulation assumes a homogeneous Poisson population, an idealization that abstracts away richer neural dynamics. In particular, inhomogeneous Poisson processes are more biologically realistic but also invalidate the property that cumulative spike counts are sufficient statistics for posterior inference. Addressing this would likely require more expressive decoding architectures, for example, incorporating recurrence or state-space formulations. 

Third, the present architecture is restricted to perceptual decisions on static images and thus cannot capture tasks with temporally varying stimuli (e.g., the random-dot motion paradigm). Extending the model to handle dynamic evidence would broaden its scope to a wider range of decision-making contexts and allow comparison to several classic experiments connecting neural and behavioral data.

\textbf{Conclusion. }
Our findings reveal that image-computable models can capture key dynamics of perceptual decision making by linking neural activity to choices and response times. This suggests a promising direction toward more expressive models of biological neural computation.

\section*{Acknowledgments and Funding} We thank Thomas Christie for his thoughtful comments on the manuscript and for the conversations that helped shape and develop this line of work. We also thank Jonathan Pillow, Chris Summerfield, Fabian Sinz, Srini Turaga, Leyla Isik, and Janne Lappalainen for their valuable feedback and discussions during the early development of this project at the Cajal NeuroAI summer course. Hayden R. Johnson was supported by
an FWO grant (G053624N). Anastasia N. Krouglova was supported by an FWO grant (G097022N).

\printbibliography

\newpage
\appendix
\onecolumn

\appendix
\section{Proof of convergence}
\label{app:proof-cond-mle}

Let \(\{(a_i, \rvz_{t,i})\}_{i=1}^N\) be i.i.d.\ samples from the true joint
\(p(a)\,p(\rvz_t\mid a)\), where \(p(a)\) is the (fixed) prior over \(a\) and
\(p(\rvz_t\mid a)\) is the likelihood of \(a\).
We fit a conditional model \(p_\theta(a\mid \rvz_t)\) by maximum likelihood.

Maximizing the product \(\prod_{i=1}^N p_\theta(a_i\mid \rvz_{t,i})\) with respect to \(\theta\) is equivalent to maximizing the average log-likelihood
\begin{equation}
\label{eq:emp-avg}
\frac{1}{N}\sum_{i=1}^N \log p_\theta(a_i\mid \rvz_{t,i}).
\end{equation}
By the strong law of large numbers, as \(N\to\infty\),
\begin{equation}
\label{eq:slln}
\frac{1}{N}\sum_{i=1}^N \log p_\theta(a_i\mid \rvz_{t,i})
\;\xrightarrow{\text{a.s.}}\;
\mathbb{E}_{p(a)\,p(z_t\mid a)}\!\left[\log p_\theta(a\mid \rvz_t)\right].
\end{equation}
Let \(p(\rvz_t)=\int p(a)\,p(\rvz_t\mid a)\,da\) be the induced marginal of \(\rvz_t\).
Consider the KL divergence between the two joint distributions on \((a,\rvz_t)\):
\begin{equation}
\label{eq:kl}
D_{\mathrm{KL}}\!\Big(p(a)\,p(\rvz_t\mid a)\,\Big\|\,p(\rvz_t)\,p_\theta(a\mid \rvz_t)\Big)
=
\mathbb{E}_{p(a)\,p(\rvz_t\mid a)}
\!\left[
\log\frac{p(a)\,p(\rvz_t\mid a)}{p(\rvz_t)\,p_\theta(a\mid \rvz_t)}
\right].
\end{equation}
Rearranging \eqref{eq:kl} yields
\begin{equation}
\label{eq:exp-ll-equals-kl}
\mathbb{E}_{p(a)\,p(\rvz_t\mid a)}\!\left[\log p_\theta(a\mid \rvz_t)\right]
=
-\,D_{\mathrm{KL}}\!\Big(p(a)\,p(\rvz_t\mid a)\,\Big\|\,p(\rvz_t)\,p_\theta(a\mid \rvz_t)\Big)
+ \text{const.},
\end{equation}
where the constant does not depend on \(\theta\).
Thus, maximizing the expected log-likelihood is equivalent to minimizing the KL divergence in \eqref{eq:kl}.

The KL divergence in \eqref{eq:kl} is minimized (to \(0\)) if and only if the two joint distributions are equal almost everywhere:
\begin{equation}
\label{eq:joint-equality}
p(a)\,p(\rvz_t\mid a) \;=\; p(z_t)\,p_\theta(a\mid z_t)\qquad\text{a.e.}
\end{equation}
Solving \eqref{eq:joint-equality} for \(p_\theta(a\mid \rvz_t)\) gives
\begin{equation}
\label{eq:posterior}
p_\theta(a\mid z_t) \;=\; \frac{p(a)\,p(\rvz_t\mid a)}{p(\rvz_t)}
\;=\; p(a\mid \rvz_t).
\end{equation}
Combining \eqref{eq:slln} and \eqref{eq:exp-ll-equals-kl}--\eqref{eq:posterior}, we conclude that any maximizer \(\theta^\star\) of the limiting objective satisfies
\(p_{\theta^\star}(a\mid \rvz_t)=p(a\mid \rvz_t)\) almost everywhere. Consequently, under standard regularity (realizability of the true posterior within the model class and optimization that attains the global maximum), the MLE estimator satisfies
\[
p_{\hat\theta_N}(a\mid \rvz_t)\;\longrightarrow\;
p(a\mid \rvz_t)
\quad\text{as } N\to\infty.
\]
\hfill\(\square\)

\section{Varying the latent dimension}
\label{app:varying-latent}
The latent dimensionality is a core architectural hyperparameter that is fixed prior to training and constrains the model’s representational capacity. In our main experiments, we set the latent dimensionality to 128, following \textcite{vafaii2024poisson}. Here, we systematically vary this dimensionality to examine its influence on three aspects of model behavior, which we investigate in turn:

\begin{center}
\begin{minipage}{0.6\linewidth}
\begin{mdframed}[linewidth=0.7pt,
                 innerleftmargin=6pt,
                 innerrightmargin=6pt,
                 innertopmargin=4pt,
                 innerbottommargin=4pt]
\begin{enumerate}[label=(\roman*)]
    \item the quality of \pvae{} reconstructions,
    \item the rate of evidence accumulation, and
    \item the sparsity of the learned latent representation.
\end{enumerate}
\end{mdframed}
\end{minipage}
\end{center}

\subsection{Reconstruction of the MNIST dataset}
\begin{figure}[h!]
    \centering
    \includegraphics[width=0.9\linewidth]{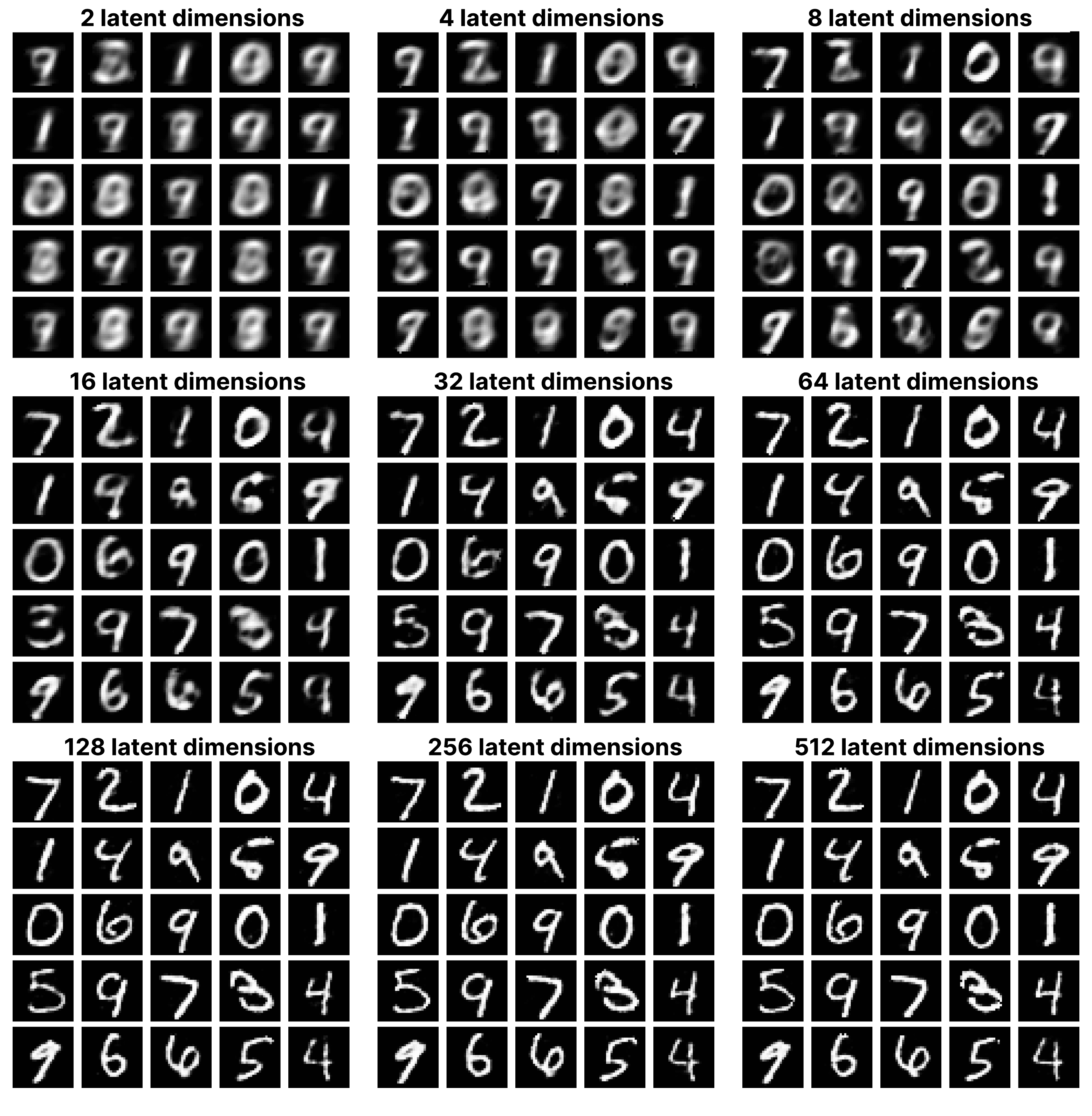}
    \caption{\textbf{Reconstruction quality of the MNIST dataset.} The reconstruction of 25 MNIST images from the test set for \pvae with varying latent dimensions. Reconstruction improves as latent dimension increases, indicating increased capacity.}
    \label{fig:app_reconstruction}
\end{figure}

To assess how latent dimensionality affects reconstruction fidelity, we varied the number of latent dimensions in \pvae on a logarithmic scale from 2 to 512, holding other training hyperparameters constant. As expected, reconstruction quality improves monotonically with dimensionality (Figure~\ref{fig:app_reconstruction}), reflecting increased representational capacity. At very low dimensions, the bottleneck severely restricts the encoder’s ability to preserve image structure, whereas higher-dimensional latents enable accurate reconstructions.

\subsection{Rate of evidence accumulation}
\begin{figure}[h]
    \centering
    \includegraphics[width=0.6\linewidth]{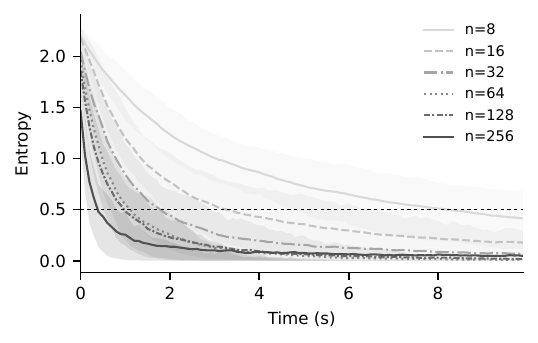}
    \caption{Mean entropy traces for $n$-dimensional latent representations in \pvae-RT across 1000 images and 100 trials. Decreasing the dimensionality of the latent reduces the rate of evidence accumulation, leading to slower response times.}
    \label{fig:mean_entropy_n_neurons}
\end{figure}

We next analyzed how latent dimensionality influences the rate of evidence accumulation. We quantified this effect using the mean posterior entropy $\mathcal{H}[p_\theta]$ across images and trials in \pvae-RT. Posterior entropy decreases stochastically within trials and, when averaged, yields smooth traces characterizing the accumulation rate. Consistent with \textcite{christie2023information}, higher latent dimensionality accelerates entropy reduction -- interpreted as increased signal power -- while lower-dimensional spaces constrain evidence accumulation, producing slower response times (Figure~\ref{fig:mean_entropy_n_neurons}).

\subsection{Sparsity of learned codes}
\begin{figure}[H]
    \centering
    \begin{minipage}{0.45\linewidth}
        \centering
        \includegraphics[width=\linewidth]{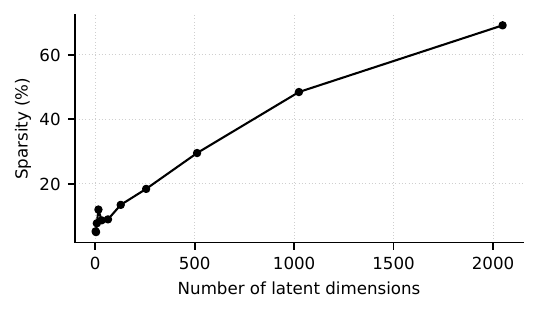}
    \end{minipage}\hfill
    \begin{minipage}{0.45\linewidth}
        \centering
        \includegraphics[width=\linewidth]{ 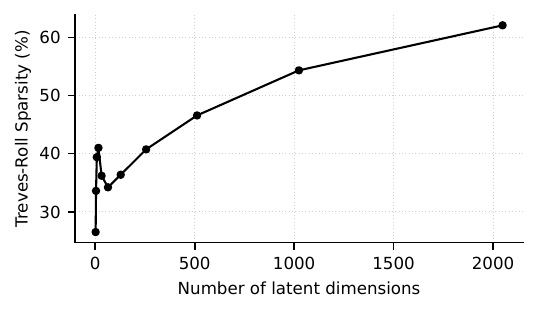}
    \end{minipage}
    \caption{\textbf{Sparsity over latent dimensions.} (\textit{Left}) Sparsity as a proportion of neurons exceeding threshold $\Psi=1$. (\textit{Right}) Treves-Rolls Sparseness index.}
    \label{fig:percent_neurons}
\end{figure}

Inspired by the relation to sparse coding, we examine the sparsity of the learned representation as a function of latent dimensionality. First, sparsity was quantified as the proportion of neurons whose responses exceeded a threshold (over 1000 images):
\begin{equation}
    \text{sparsity \%} = \frac{\#\{\text{neurons with response > }\Psi\}}{\text{total neurons}} \times 100,
\end{equation}
where $\Psi$ is a tunable threshold. Across thresholds, sparsity consistently increased with latent dimensionality -- an example for $\Psi = 1$ is shown in Figure~\ref{fig:percent_neurons} (left).

To validate our results, we additionally quantified the lifetime sparsity using the Treves–Rolls sparseness index \cite{rolls1995sparseness}, which captures the distributional shape of firing rates $r$:
\begin{equation}
    \alpha=\frac{\left(\frac{1}{N} \sum_i r_i\right)^2}{\frac{1}{N} \sum_i r_i^2}, 
   \qquad \text{TR sparsity \%} = \frac{1 - \alpha}{1-1/N},
\end{equation}
where $N$ is the number of images and $r_i$ denotes the mean firing rate of a neuron in response to image $i$.

Following the analysis in \textcite{christie2023information}, we varied the number of latent dimensions (i.e., neurons) from 2 to 2048 on a logarithmic scale. Both metrics revealed the same overall trend: sparsity increases with latent dimensionality. 

\newpage

\section{Training details}
\label{app:training}
All training can be completed in a few hours under modest compute resources.

\subsection{Software and hardware}

Programming language: Python 3.10. Core ML package: PyTorch 2.5.1. Analysis packages: NumPy, Matplotlib, Pandas, scikit-learn. Hardware: 2024 MacBook Pro with Apple M4 Max.

\subsection{Poisson VAE}
\textbf{Dataset}: MNIST; 28×28 grayscale images flattened.

\textbf{Architecture}:\\ 
Latent dimensionality: 128 \\
Encoder, $\text{enc}_\phi(x)$: 784 → 128 (linear). \\
Decoder, $\text{dec}_\psi(z)$: 128 → 784 (linear) with Gaussian output.

\textbf{Training}: ELBO objective; Adam optimizer with learning rate 1e-3; 50 epochs.


\subsection{Task-optimized decoder}
\textbf{Spike processing:} For each stimulus, we generate spike trains from the image-driven rates 
$\lambda = \mathrm{enc}_\phi(x)$ using independent homogeneous Poisson processes. 
The spike trains are discretized into $T=100$ time bins to form a binary event matrix 
(neurons $\times$ time). This is further converted into cumulative spike counts, where 
entry $(i,t)$ denotes the total number of spikes emitted by neuron $i$ up to time $t$. 
Each cumulative count vector $\zz_t$, paired with its label $a$, serves as a supervised 
training sample.

\textbf{Architecture:} multilayer perceptron 128 → 64 → 32 → output size; ReLU activations in hidden layers; final layer produces class logits.

\textbf{Training:} cross-entropy loss; Adam optimizer with learning rate 1e-3; batch size 256; 100 epochs.

\end{document}